\documentstyle[prl,aps,epsf,multicol]{revtex}
\begin{document}
\draft

\title{Phase separation driven by
a fluctuating two-dimensional self-affine potential field}

\author{G. Manoj$^{1,2}$ and Mustansir Barma$^{3}$}

\address{$^{1}$ The Institute of Mathematical Sciences, C. I. T. Campus, Chennai 600 113, India.\\
$^{2}$ Department of Physics, Virginia Polytechnic Institute and State University, Blacksburg, VA 24061-0435, USA.\thanks{Present Address.}\\
$^{3}$ Department of Theoretical Physics, Tata Institute of Fundamental
Research, Mumbai 400 005, India.}

\date{\today}

\maketitle
\begin{abstract}

We study phase separation in a system of hard-core particles driven by
a fluctuating two-dimensional self-affine potential landscape which
evolves through Kardar-Parisi-Zhang (KPZ) dynamics.  We find that
particles tend to cluster together on a length scale which grows in
time.  The final phase-separated steady state is characterized by an
unusual cusp singularity in the scaled correlation function and a
broad distribution for the order parameter. Unlike the one-dimensional
case studied earlier, the cluster-size distribution is asymmetric
between particles and holes, reflecting the broken reflection symmetry
of the KPZ dynamics, and has a contribution from an infinite
cluster in addition to a power law part.  A study of the surface in terms
of coarse-grained depth variables helps understand many of these
features.

\end{abstract}

\vspace{1cm}
\begin{multicols}{2}

\section{INTRODUCTION}

The behaviour of a scalar field driven by a fluctuating force
field depends strongly on the correlations of the driving field in space
and time. A well known example is the passive scalar problem 
in fluid mechanics, where one asks for the behavior of a passive field
as it is advected by a turbulent fluid flow \cite{KRAICH}.
Even in situations where the driving force field has simpler
correlations, the passive scalar field can show interesting, and
sometimes unexpected, behaviour. 
In particular, while in the fluid
context an initial local concentration of passive particles typically
spreads out in space, in other types of situations an initially randomly 
distributed set
of particles may be driven into a state with large-scale clustering.
In this paper, we study one such example.

We consider a force field which is derived from a fluctuating potential,
and ask for its effect on a system of particles which do not interact
with each other except through hard-core exclusion. The problem is then
tantamount to the dynamics of hard-core particles which reside on a
fluctuating surface and are driven downwards by gravity along local
slopes.  An especially interesting case arises when the surface is
self-affine with a power-law divergence in the height correlation
function as a function of the separation. In such cases, surface
roughening strongly affects the clustering of particles
and can lead to new types of states\cite{KARD,DAS}. 
We study a stochastically evolving two-dimensional surface governed by
Kardar-Parisi-Zhang (KPZ) dynamics \cite{KPZ}
and show that surface fluctuations bring about large-scale clustering
of particles, akin to phase separation.  The particles are taken to be
random walkers with an excluded volume constraint, diffusing in 
the dynamic potential landscape defined by the local height $h({\bf r},t)$
at a base position ${\bf r} \equiv (r_x,r_y)$
 at time $t$. In addition, the particles are assumed to be
sufficiently massive that the dynamics operates effectively at zero temperature;
this means that particles only move locally downwards at a fixed rate, subject 
to the conditions that the local slope is favourable and the target site is
unoccupied. This model is a generalization to two dimensions of the
1-d model studied in
\cite{DAS} where it was found that the particles reach a phase
separated state with unusual properties arising from strong surface
fluctuations. Since the nature and strength of fluctuations depends
strongly on the dimensionality, it is important to see to what extent
these features survive in higher dimensions. This is one of the
principal aims of this paper.

Let us summarize the main results. Our numerical simulations support 
the argument that the particle density exhibits phase ordering over a
characteristic time-dependent coarsening length scale 
${\cal L}(t)\sim t^{1/z}$ set
by surface fluctuations \cite{DAS}.  Here $z$ is the dynamical exponent for
surface fluctuations; for the KPZ surface with $d=2$, it is known that
$z \simeq 1.6$ \cite{MOORE}.
The pair-correlation function is found to depend on the separation scaled by
${\cal L}(t)$, as is characteristic of a system undergoing
phase ordering, but the scaling
function has a cusp near the origin, unlike usual phase separating
systems \cite{BRAY}.  The cluster size distribution in the steady state has a
power law decay $N(s)\sim s^{-\tau}$ for $s\ll L^{2}$, where the
exponent $\tau$ is different for particles and holes. In addition, unlike
the 1-d case,
there is a distinct contribution to $N(s)$ from an `infinite' cluster
which contains a finite fraction of the total number of particles.  An
understanding of these results can be gained from a study of the
surface itself.  To this end, it proves useful to distinguish between regions
where the height is less than or more than a fixed reference level. This
is incorporated in a Coarse-grained Depth (CD) model of the surface 
\cite{DAS,KIM}, which itself exhibits phase
separation. Many of its properties reflect the underlying
asymmetry of the KPZ surface. Finally, in both the sliding particle and depth
models, the order parameter has a distribution that 
remains broad in the thermodynamic limit. 

We introduce the model in Section II. In Section III, we present
results for two-point correlation functions, cluster-size distributions
and the order-parameter distribution, while Section IV is the conclusion.

\section{MODEL}

We study an autonomously evolving two-dimensional surface, whose
behaviour over large length and time scales is described by the
Kardar-Parisi-Zhang equation\cite{KPZ}

\begin{equation}
\frac{\partial h}{\partial t}=\nu \nabla^{2}h+\lambda (\nabla h)^{2}+\eta({\bf r},t)
\end{equation}
with short range correlations for the noise, $\langle \eta({\bf r},t)
\eta({\bf r}^{\prime},t^{\prime})\rangle \propto
\delta^{d}({\bf r}-{\bf r}^{\prime})\delta(t-t^{\prime})$.  
The KPZ surface in $d=2$ is known to be self-affine in steady state:
$\langle (h(0) - h({\bf r}))^2\rangle \sim r^{2\chi}$ where $\chi$ is
the roughness exponent.  For the KPZ problem with $d=2$, $\chi \simeq
0.4$\cite{MOORE,LASSIG}.  The nonlinear term in Eq.(1) breaks $h
\rightarrow -h$ symmetry, and this has important consequences for the
models we study. 

We simulate the surface through a discrete solid-on-solid (SOS)
algorithm\cite{MEAKIN}, where the height difference between nearest
neighbour (NN) points on a square lattice is maintained at $\pm 1$. A
point is selected at random and its height is increased by 2 units
with probability $p_{+}$ if all four of its NN points are at greater
height, and decreased with probability $p_{-}$ if all four are at a
lower height. Otherwise, the site is not updated. It is believed that
the asymptotic properties of this single step model are the same as
those of the (2+1)-dimensional KPZ equation, though this has not been
proved.  In our simulations we chose $p_{+}=0$ and $p_{-}=1$, so
that we have an `evaporating', rather than a growing surface whose
average height decreases in time.

The Sliding Particle (SP) model is defined as follows.  Particles are initially
distributed at random on surface sites with no more than one particle
per site, the overall particle density being $\rho$. The external
force field which drives the particles acts downwards, so that particles
tend to move in the same direction as the average surface height, 
while holes tend to move upwards, opposite to the direction of surface
motion. In a microstep, a randomly chosen
particle attempts a move to a randomly chosen neighbouring site. If the local
slope is favourable and the target site is unoccupied, the move is
made, otherwise the particle stays at the same site. One MC step is
counted when, on average, all surface sites and all particles have been 
updated in
a random sequence. We verified that a faster rate of updating for
particles with respect to surface updates does not change the
qualitative results. We did our simulations on square lattices with
$\rho = \frac {1}{2}$ and sizes ranging from $L=32$ to $L=256$.  We
used periodic boundary conditions for both the surface and
the particles which reside on it.

It is helpful to introduce two sets of discrete Ising-like spin
variables to characterise the particle and surface configurations. For
the particle configuration, we define $\sigma({\bf r})=2n({\bf r})-1$
where $n({\bf r})$ is the local occupation index, i.e., $n({\bf r})=1$
if there is a particle at ${\bf r}$ and zero otherwise.  Further, to
characterise the height fluctuations of the surface configuration, we
define a Coarse-grained Depth model in which we categorize sites according
to whether they are above or below a certain fixed height.  To this
end, we define $s({\bf r})=-sgn(h({\bf r})-h_{0})$ where $h_{0}$ is a
chosen reference level. We call the set $\{\sigma({\bf r})\}$ as SP
spins and the set $\{s({\bf r})\}$ as CD spins. In order to have a rough
correspondence with
the half-filled case $\rho = \frac {1}{2}$ of the SP model, we take
$h_{0}$ to be a spatial
average over the configuration of surface heights at that instant.

We expect a correlation between SP and CD spins as local slopes in the
surface guide the particles towards the local minima, so that over
sufficiently large time scales, particles are expected to
preferentially cluster in
low-depth regions with predominantly positive CD spins. If particles
were to occupy the lowest available positions so as to minimise the
total potential energy, we would have a close match of the $\sigma({\bf r})$
and $s({\bf r})$. However, this `ground state' is never
reached as the surface reconfigures itself before particle
rearrangements can occur.

\section{RESULTS}
\subsection{Correlation Functions}

The SP-CD correspondence gives an insight into the extent of
clustering of the particles. For example, a hill (a region with
negative $s(r)$) of linear extension $\xi$ is expected to
overturn in typical time $\tau\sim \xi^{z}$, causing clustering of
particles as they fall into the valleys that form.
  Since the largest hill that
overturns within time $t$ has typical size ${\cal L}(t)\sim t^{1/z}$, this
defines the characteristic length scale for particle clustering
\cite{DAS}. This
is verified in a numerical study of the pair spin correlation function
$C(r,t)=\langle \sigma({\bf 0},t)\sigma({\bf r},t)\rangle$ in the SP
model. As we see in Fig.1, this function can be fitted into a scaling
form $C(r,t)=f(r/{\cal L}(t))$ with ${\cal L}(t)\sim t^{1/z}$
and $z=1.6$, which is close to
the value of the dynamical exponent for the (2+1) dimensional KPZ
surface \cite{MOORE}. This dynamic scaling form is characteristic of a
phase ordering system. However, the scaling function has a cusp
singularity at small argument, unlike usual
coarsening systems where the function is linear for small
arguments \cite{BRAY}. 

\begin{figure}
\epsfxsize=1.6in
\hspace{-1.0cm}
\epsfbox{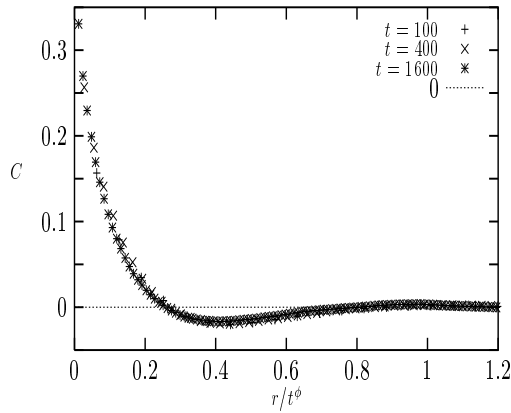}
\narrowtext
\caption{ Equal-time two-point spin correlation for particles sliding
down a 2-d KPZ surface
of linear size 256. There is a scaling collapse when plotted as a function
of the scaled variable $r/t^{\phi}$ with $\phi=0.6$, indicating  coarsening. 
Each point represents an average over 100 different time histories.}
\end{figure}

In steady state, the system size $L$ replaces ${\cal L} (t)$ as the relevant
length scale, and a similar scaling form is expected
for the correlation function.
To quantify the deviation from linearity at small arguments, we
studied the structure factor in the steady state,

\begin{equation}
S({\bf k},L)\equiv \int d^{2}{\bf r}C({\bf r},L)e^{-i{\bf k.r}}.
\label{eq:EQ2}
\end{equation}

As shown in Fig. 2, the direction-averaged structure factor is well described
by the form $S(k,L)=1-c_{0}+L^{2}g(kL)$, where $c_{0}\simeq
0.4$ and $g(q)\sim q^{-(2+\alpha)}$ at large $q$. The first term 
arises from
a short-distance analytic contribution $(1-c_0)\delta({\bf r})$
which adds on to the scaling part  $f(r/L)$ of $C({\bf r},L)$.
We find $\alpha\simeq 0.38(4)$, a pronounced difference from typical coarsening
systems where $\alpha=1$ (Porod Law) \cite{POROD}. The non-Porod form implies a
cusp in the real space scaling function at small argument,

\begin{equation}
f(x)\simeq c_{0}-c_{1}x^{\alpha}+...\hspace{0.3cm}\hspace{0.3cm} x\ll 1
\label{eq:EQ3}
\end{equation}
with $x=r/L$.

We have also studied the two-point correlation function in the CD
model, and find a similar behaviour, with a cusp exponent 
$\alpha \simeq 0.43(3)$
In the next subsection we will show that within the independent interval 
approximation, the cusp exponent $\alpha$ of the CD model is equal to the 
roughness exponent $\chi$.  

\begin{figure}
\epsfxsize=1.6in
\epsfbox{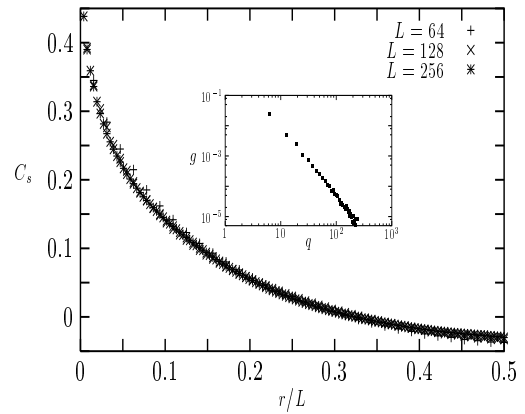}
\narrowtext
\caption{ Steady state pair correlation in the SP model for various 
system sizes $L$.
The finite size scaling form $C(r,L)=f(r/L)$ illustrated by
the data collapse is is characteristic of phase separation. Inset:
The scaled structure factor $g(q)$ plotted
against the scaled wave vector $q=kL$ with $L=128$
shows that $g(q)\sim q^{-(2+\alpha)}$ with $\alpha\simeq 0.38(4)$, implying
a cusp singularity in the real space correlation function.} 
\end{figure}

\subsection{Cluster Size Distributions}

Let us define a cluster as a set of like SP
spins (particles or holes), each 
of which is a nearest neighbor of at least one other like spin in that cluster.
A study of the size distribution of clusters shows (Fig. 3) that
the number of connected
clusters $N_{\pm}(s)$ with $s$ particles (holes) has a
power-law decay for $s \ll L^2$:

\begin{equation}
N_{\pm}(s)\sim L^{2}s^{-\tau_{\pm}}\hspace{0.3cm};\hspace{0.3cm}s\ll L^{2}
\label{eq:EQ4} 
\end{equation}

where $\tau_{+}\simeq 2.2$ and $\tau_{-}\simeq 2.0$. This power-law
distribution of cluster sizes, reminiscent of critical systems, is
another characteristic feature of the unusual phase separated state
under study ---  it occurs in one dimension as well \cite{DAS}.  
There are, however,
two important differences from the 1-d case. First, there is a marked
difference in the powers $\tau_+$ and $\tau_-$ for particle $(\sigma_i
= +1)$ and hole $(\sigma_i = -1)$ clusters, which reflects the
asymmetry of the $s_i = +1$ and $s_i = -1$ clusters in the CD model as
discussed below.  Second, for the particle cluster distribution we
find that there is an additional large $s$ contribution of the form
$f_\infty (s/L^2 - y_0)$ peaked at $y_0 \simeq 0.4$.  For finite
system sizes, $f_\infty$ is somewhat broad, but with increasing $L$,
the width of the peak narrows down.
We find that the area under the peak is unity, implying that there is
a single very large cluster in every configuration. Also, the
fraction of particles contained in this cluster is $2 y_0 \simeq 0.8$.

The power-law distribution of cluster sizes in the SP model has its counterpart in the CD model too.
It is known that when a rough surface is intersected by a 
horizontal plane, 
the distribution of areas enclosed by the closed contours of intersection 
has the power-law form
${\cal N}(s)\sim s^{-\tau^{*}}$ for $s\ll L^{2}$, where 
$\tau^{*}=2-\frac{\chi}{2}$
\cite{MATSUSHITA,ZEITAK,KONDEV}.  A typical contour encloses
several other contours, and the overall structure is scale 
invariant\cite{MEAKIN}.
For the KPZ surface one should moreover distinguish between contours
whose inner perimeter sites lie above or below the cut, as
there is no {\it a priori} reason for both the distributions to be
identical. In fact, numerical simulations
show that cluster size distributions for CD spins (which are
closely related to area distributions), indeed
has an asymmetry between positive and negative spins (Fig. 4). Both 
the distributions 
follow power-law decays, but
with different exponents. The exponent for negative CD spins is numerically 
close to $\tau^{*}$, while
that for positive spins is significantly larger. Morover, there is a 
contribution from an infinite cluster of positive spins,
as for the particle distribution in the SP problem. 

\begin{figure}
\epsfxsize=1.8in
\epsfbox{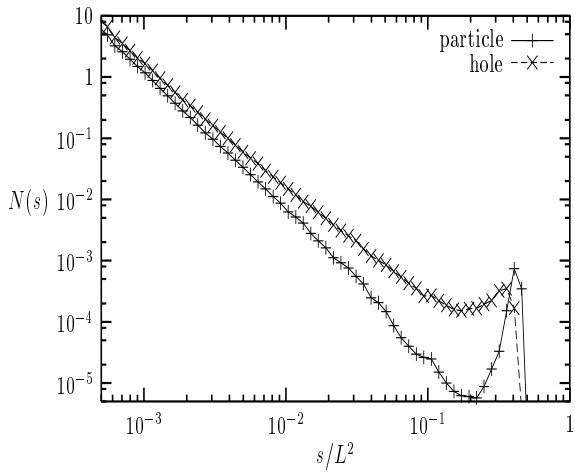}
\narrowtext
\caption{ Size distribution of cluster sizes 
for $+$ spins (particles) and $-$ spins (holes) for the SP model.
The data shows a power-law decay with
exponent $\simeq 2.2$ and a well-separated `infinite' cluster
for particles, and a lower value of the exponent
$\simeq 2.0$ for holes. We used $L=128$ and averaged over 100 histories.}
\end{figure}

Within the CD model, we argue that
power laws in cluster distributions of a different sort are related
to the occurrence of a cusp in the scaled correlation function.
For a self-affine surface with
roughness exponent $\chi$, it is known that the probability
$P(l)$ that the surface first returns to its starting height $h({\bf
x=0})=h_{0}$ after moving a distance $l$ along an arbitrary linear
direction has a power-law decay at small $l$: $P(l)\sim l^{-(2-\chi)}$
for $l\ll L$\cite{KIM,MATSUSHITA,SHMITT}. Each such segment defines a linear cluster of
CD spins $s_i$ of the same sign, along the linear cut.
Now let us make the independent interval approximation (IIA) in which the 
lengths of such segments are taken to be independent random variables. 
This enables the Laplace transforms of the cluster size distribution and
pair correlation function for CD spins (defined relative to $h_{0}$)
to be related along any linear cut. The nonstandard feature is that the mean 
cluster size diverges as $L \rightarrow \infty$, but the correlation function
can be calculated as in \cite{DAS}. It has the
scaling form
\begin{equation}
C^{*}(r,L)\simeq 1-a(\frac{r}{L})^{\chi}+..\hspace{0.3cm};\hspace{0.3cm}r\ll L
~~~~(IIA)
\label{eq:EQ5}
\end{equation}
Comparing with Eq.\ref{eq:EQ3}, we see that the IIA predicts that
the cusp exponent $\alpha$
in the CD model is equal to the roughness exponent $\chi$. The KPZ value 
$\chi$ is quite close to the measured value of $\alpha$ for the CD model,
and also to that for the SP model.  The behaviours of both models on large 
scales of distance and time appear to be similar in 2-d, even though the 
microscopic configurations
of the two match only roughly. To get an idea of the latter, we 
monitored the overlap index $O=\langle \sigma({\bf r})s({\bf r})\rangle$, 
and found that $O$ varies between $\simeq 0.38$ when the frequency of updates 
of particles and surface is equal, and a saturation value $\simeq 0.6$ 
as the ratio between the two is increased. However, there was no 
corresponding significant
change in the numerical values of the exponents $\alpha$ and $\tau$.

\begin{figure}
\epsfxsize=1.8in
\hspace{-2.0cm}
\epsfbox{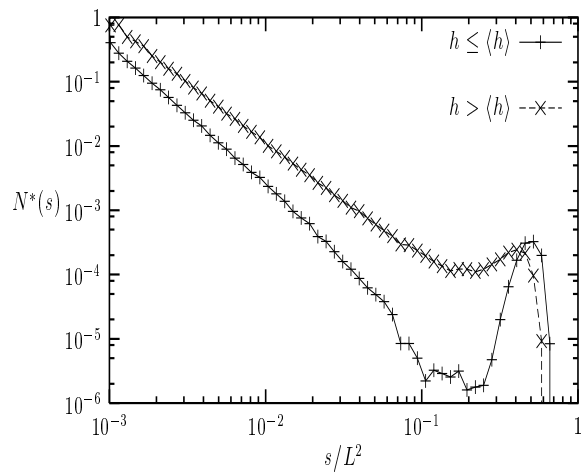}
\narrowtext
\caption{ Domain size distribution for $+$ (valley)  and $-$(hill) spins 
for the CD model. The exponents for the power-law parts
have values $\simeq 2.2$ and
1.85 respectively. There is a distinct contribution 
to the distribution of $+$ spins from an infinite cluster.
We used $L=128$ and averaged over 100 histories.}
\end{figure}

\subsection{Order Parameter}

For a system with conserved magnetisation like the SP model, an appropriate
quantity to characterise the ordered state is the steady state average of the 
magnitude of the Fourier components
of the density \cite{DAS}, defined as $Q({\bf k})=
\langle |L^{-2}\sum_{\bf r}n({\bf r})e^{-i{\bf k.r}}|\rangle$
where ${\bf k}=\frac{2\pi}{L}(n_{x},n_{y})$ and $|{\bf k}|\leq \pi$.
Taking the magnitude  guarantees that $Q(\bf k)$ receives the same
contribution from all configurations that can be obtained from each other
by translational shifts \cite{SCHMITT}.
In Fig. 5, we plot $Q({\bf k})$ along the (1,0) direction for four
different lattice 
sizes. The sequence of curves suggests that
for any fixed, finite $k$, $Q(k=2\pi n/L,0)$ approaches zero as $L$ increases.
However, if $n$ is held fixed so that $|{\bf k}|$ approaches zero as $L
\rightarrow \infty$, the corresponding $Q$ approaches a finite limit.
Of this set of $Q$'s, the largest is $Q^{*}=Q(\frac{2\pi}{L},0)$, and
provides the simplest characterization of the order. 

\begin{figure}
\epsfxsize=1.8in
\hspace{-1.0cm}
\epsfbox{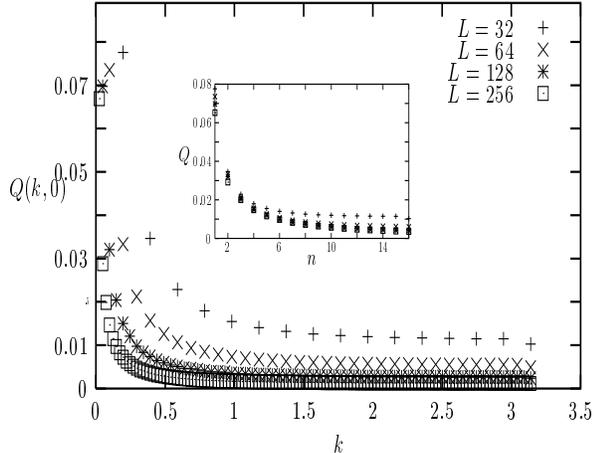}
\narrowtext
\caption{ $Q(k,0)$ plotted against wave vector $k =2 \pi n/L$ and
(Inset)$n$ for four lattice sizes.}
\end{figure}

\begin{figure}
\epsfxsize=1.8in
\hspace{-1.0cm}
\epsfbox{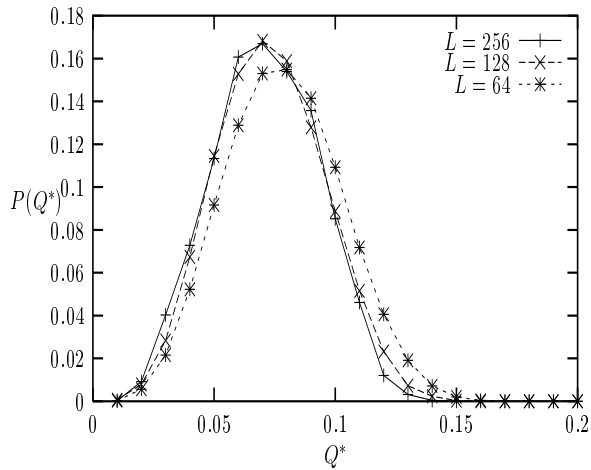}
\narrowtext
\caption{ Probability distribution of the order parameter $Q^{*}$ for three
lattice sizes for the SP problem.
}
\end{figure}

Fluctuations in the SP model are unusually strong and are reflected in
the probability distribution of $Q^{*}$ (Fig. 6).
The distribution seems to remain broad and approach a limit as
$L$ increases, with $\langle Q^{*}\rangle\simeq
0.07$ and variance $\simeq 0.02$. Similar studies of the corresponding 
CD model shows
that there is a broad distribution there as well, 
and we find $\langle Q^{*}\rangle\simeq
0.11$. The fact that the RMS fluctuation of the order parameter does not
vanish in the large-size limit is an unusual and characteristic feature of
the ordered state of this model.  Note that
fluctuations which drive $Q^{*}$ towards zero need not take the system into
a disordered state. A study of the temporal fluctuations of $Q^{*}$ in one
dimension showed that a low value of $Q^{*}$ occurs simultaneosly
with an increase of the value of 
$Q$ for another close-by  $n$, so that the state 
retains macroscopic order, but with a few more coexisting macroscopic domains
\cite{DAS}.  A more complete characterization of the order in 
two dimensions would thus involve finding the joint probability ${\cal P}
[ Q(\frac{2\pi n_{x}}{L},\frac{2 \pi n_{y}}{L})]$ as $L \rightarrow \infty$,
but this has not been attempted here. 

It is interesting to compare the distribution $P(Q^{*})$ for this
model with the corresponding quantity for a more familiar system such as
the 2-d Ising model evolving under conserved (Kawasaki) dynamics.  For
the ferromagnetic Ising model with equal numbers of up and down spins, 
the ordered state in a finite system
consists of strips of width $ \sim {1\over 2} L$ parallel to either $x$
or $y$ axes, with equal probability.  Thus the distribution $P(Q^{*})$
(say, with $Q^{*}=Q(\frac{2\pi}{L},0)$ ) in that case would have two
peaks, one at a non-zero value of $Q^{*}$ corresponding to strips
forming along the $x$-direction, and another peak at $Q^{*}=0$ which
corresponds to strips forming in the $y$-direction. In our case,
however, the peak at the origin is absent as the interface between phases is
much more diffuse and there is no strip formation.  In this sense, the ordering
observed in this model is very different from that in traditional
equilibrium models.

\section{CONCLUSION}

To conclude, we have studied a model where a fluctuating self-affine
surface drives a set of downward-drifting particles to a phase
separated state. The phase separation has several unusual
characteristics arising from the presence of strong fluctuations which
survive even in the thermodynamic limit.  These include a power-law
distribution of cluster sizes, a cusp in the scaled pair correlation
function and a finite width of the order parameter distribution. These
features mirror the properties of a coarse-grained height model of
surface fluctuations, where they follow directly from the self-affine
nature of surface fluctuations. In particular, the asymmetry in the
cluster size distribution for particles and holes is a consequence of
the fact that KPZ growth breaks up-down symmetry. It would be
interesting to see if a similar asymmetry exists in other related
quantities such as the distribution of lengths of contours and the
areas enclosed by them.

It would also be interesting to study the sliding particle problem on
a fluctuating Edwards-Wilkinson surface, where particle-hole symmetry
should be restored in the half-filled case. This surface is
logarithmically rough in 2-d, suggesting that the scaled two-point
correlation functions for the sliding particle and depth problems
should show an even sharper cusp than for the KPZ surface studied in
this paper.

Finally, we remark that it would be interesting to explore the effects
of removing the hard core interaction between particles. A study of
noninteracting particles sliding on a KPZ surface in one dimension
indicates a power law decay of the two-point correlation function
\cite{KARD}. A study of other characteristics of the resulting state
should prove quite revealing.

\section{ACKNOWLEDGEMENTS}

We would like to thank D. Das, D. Dhar, S.N. Majumdar, B. Schmittmann and
R.K.P Zia for helpful discussions. G. M. acknowledges the hospitality of 
the Tata Institute of Fundamental Research where this work was initiated. 
This research is supported in part by a grant (DMR-0088451) from the
US National Science Foundation.

It is a pleasure and a privilege to contribute this article to the
Festschrift volume for the 70th birthday of Prof. Michael E. Fisher,
whose work on various aspects of co-operative ordering has inspired
several generations of statistical physicists. MB would like to
acknowledge many stimulating interactions over the years with
Professor Fisher, both professional and personal.

\end{multicols}
\end{document}